\documentclass[12pt,a4paper]{article}
\usepackage{amsmath, amssymb}
\usepackage{slashed}
\usepackage{verbatim}
\usepackage{latexsym}
\usepackage{euscript}
\usepackage{amsfonts}
\usepackage{verbatim}
\usepackage[]{array}
\usepackage[]{mathrsfs}
\usepackage{graphicx}
\usepackage{amsmath}
\usepackage{amssymb}
\usepackage{amsxtra}
\usepackage{color}
\usepackage{ulem} 
\def\be{\begin{eqnarray}}
\def\ee{\end{eqnarray}}
\def\0{\nonumber}

\def\det{\rm det}
\def\bfh{{\bf h}}

\def\bfd{{\bf d}}
\def\bfD{{\bf D}}
\usepackage{slashed}
\usepackage{verbatim}
\usepackage{latexsym}\usepackage{color}
\usepackage{euscript}

\newcommand\EK{\EuScript{K}}
\newcommand\ER{\EuScript{R}}
\newcommand\EP{\EuScript{P}}
\newcommand\EA{\EuScript{A}}
\newcommand\ED{\EuScript{D}}
\normalfont\large
\begin{document}
\begin{flushright}
SISSA/26/2020/FISI
\end{flushright}
\vskip 2cm
\begin{center}
{\LARGE {\bf HS Yang-Mills-like models: a review}}

\vskip 1cm

{\large  L.~Bonora$^{a}$, S.~Giaccari$^{b}$,\\
\textit{${}^{a}$ International School for Advanced Studies (SISSA),\\Via
Bonomea 265, 34136 Trieste, Italy, and INFN, Sezione di
Trieste }\\ 
\textit{${}^{b}$ Department of Sciences,
Holon Institute of Technology (HIT),\\
52 Golomb St., Holon 5810201, Israel}
}
\vskip 1cm

\end{center}

 \vskip 1cm {\bf Abstract.} We review the attempt to construct massless gauge field theories in Minkowski spacetime that go under the name of HS-YM. We present their actions and their symmetries. We motivate their gravitational interpretation. In particular we show how to recover the local Lorentz invariance, which is absent in the original formulation of the theories. Then we propose a perturbative quantization in the so-called frozen momentum frame. We discuss physical and unphysical modes and show how to deal with them. Finally we uncover the gauge symmetry hidden under such unphysical modes. This requires a nonlocal reformulation of the theory, which is, however, characterized by an augmented degree of symmetry.

\vskip 1cm

\begin{center}

{\tt Email: bonora@sissa.it,stefanog@hit.ac.il}

\end{center}

\vskip 1cm

\section{Introduction}

This contribution is a progress report on a, perhaps too ambitious, research program which would like to answer the following question: the unification of all the fundamental forces of nature has always been and still is a major challenge for theoretical physics; superstring theories constitute the most successful and authoritative attempt to cope with it, but a stimulating question still hovers in the air: are they the only possibility? In other words, is it not possible to construct a consistent {(i.e. unitary and renormalizable)} field theory that describes all interactions without resorting to string theory? Or, at least, a theory that bridges over the distance between (super)string theories and the two effective standard models that describe the present data on the four fundamental interactions?

We are well aware of the type of challenge we are facing with this program. There is enough evidence that a theory describing all the fundamental interactions needs an infinite number of fields of increasing spins, and, since the sought for theory should be as fundamental as possible, it should have  as few free parameters as possible; therefore we would like to start from a theory with massless fields. In fact, similar considerations have originated many attempts in the realm of higher spin (HS) gauge theories. On the one hand the free spectrum of massless HS particles is by now well-understood ~\cite{Majorana:1968zz,Dirac:1936tg,Fierz:1939zz,Pauli:1939xp,Wigner:1939cj,Fronsdal:1978rb,Francia:2002aa,Francia:2007qt,Francia:2007ee,Campoleoni:2012th}. On the other hand the construction of consistent interactions has turned out to be a much more challenging problem. Notwithstanding  the fundamental breaktrough of  the discovery of Vasiliev's equations  ~\cite{Vasiliev:1990en,Vasiliev:1992av,Vasiliev:1995dn}, progress in this direction is hindered by the lack of the relevant action, which renders quantization a remote target. In fact for HS gauge theories on a flat Minkowski space there are  severe constraints (no-go theorems), illustrated in a rather vast literature (see e.g. ~\cite{Metsaev:1991mt,Taronna:2017wbx,Roiban:2017iqg} and  \cite{Bekaert} and references within). and even on  anti-de Sitter (AdS) spaces ~\cite{Sleight:2017pcz}. In particular, in ~\cite{Ponomarev:2016lrm} a massless HS perturbatively local theory on flat background was seen to be characterized by a complex Hamiltonian and trivial S-matrix \footnote{ Interestingly, this theory (chiral HS gravity) turns out to be one-loop finite   ~\cite{Skvortsov:2020gpn}. It is also worthwhile to notice it can be smoothly deformed to an AdS background, where the S matrix is nontrivial and, by holographic correspondence, captures a subset of the correlation functions of vector models on the AdS boundary ~\cite{Skvortsov:2018uru}, as expected from ~\cite{Klebanov:2002ja, Aharony:2020omh}.}.  To be frank,  there seems to be no hope for massless local non-chiral theories and even for perturbatively local non-chiral theories (that is, for theories with infinitely many interacting fields, but with finitely many interaction terms at each derivative order). If a solution exists, it must be looked for in the realm of non-local theories. However if we decide to leave the safe ground of local field theories with a finite number of fields, the number of possibilities grows beyond control and the true issue becomes how to tackle the problem of non-locality in an organized and manageable way.

In this spirit, in \cite{HSYM} a new type of massless higher spin theories on flat Minkowski spacetime was proposed. They were dubbed HS YM-like models, because the form of their actions is inspired by Yang-Mills theories and the lowest spin field is clearly a Yang-Mills field. The fields of these theories (much in the style of string field theory) are
defined in the phase space of the Weyl-Wigner quantization, and they are written as  $h_a(x,u)$, where $u$ is the momentum conjugate to $x$. They are called  master fields. 
By expanding them in powers of the momentum  $u_\mu$ one gets an infinite series of ordinary fields of increasing spins.  From the  action  one can easily extract the equations of motion, which contain an infinite number of local terms, although the number of terms at each derivative order is finite. The main novelty of the HS YM-like models is that they admit a gravitational interpretation. This is due to the fact that the second field in the $u$ expansion of $h_a(x,u)$ has the  expected transformation properties of a frame field. The action is invariant under gauge transformations whose parameters can be expanded themselves in powers of $u$. The lowest order term in such expansion represents an Abelian gauge transformation, while the next one corresponds to diffeomorphisms. The frame-like component of $h_a$ transforms correctly under them. However a full gravitational interpretation requires also invariance under local Lorentz transformations, while the action of the HS YM-like models does not  explicitly possess it. But, this is one of the surprises, one can show that such an invariance is hidden and can be easily made manifest by redefining the action in a suitable way. In summary, the gravitational interpretation  of the HS YM-like models is fully legitimate.

Ref. \cite{HSYM} contains also a detailed proposal for perturbative quantization. The gauge freedom has to be fixed in order to extract the true physical degrees of freedom.  This turns out to be a complicated operation which leads to an overall redefinition of the action and uncovers a strong non-locality. The non-locality, though severe, is strictly tied to gauge invariance. If we suitably fix the gauge it disappears. In other words it is a gauge artifact, which makes it a controlled and hopefully manageable feature, the new non-local theory being characterized by a very large explicit symmetry.

The paper is organized as follows. In section 2 The HS-YM like models are introduced, both in the Abelian and non-Abelian case, and the gauge transformations discussed and interpreted. Section 3 is devoted to the recovery of local Lorentz invariance, section 4 to the BRST quantization. In section 5 we propose a perturbative quantization
in the so-called frozen momentum frame. In section 6 we deal with the physical and unphysical modes that make their appearance in this quantization and discuss the hidden gauge symmetry of the quantized theory and how to make it manifest. The last section contains some conclusions.

\section{An introduction to HS-YM like models}

Higher-spin Yang-Mills models are formulated in terms of master fields $h_a(x,u) $, which are local in the phase space $(x,u)$, where $[x^\mu, u_\nu]=i\hbar \delta^\mu_\nu$ ($\hbar$ will be set to the value 1). The master field is assumed to have an expansion in powers of $u$,
\be
h_a(x,u) &=& \sum_{n=0}^\infty \frac 1{n!} h_a^{\mu_1,\ldots,\mu_n}(x) u_{\mu_1}\ldots u_{\mu_n}\0\\
&=&A_a(x) +\chi_a^\mu(x) u_\mu 
+ \frac 12 b_a^{\mu\nu}(x) u_\mu u_\nu+\frac 16 c_a^{\mu\nu\lambda}(x) u_\mu
u_\nu u_\lambda + \ldots\label{haunn}
\ee
where $ h_a^{\mu_1,\ldots,\mu_n}(x)$ are ordinary tensor fields, symmetric in $\mu_1,\ldots,\mu_n$. In \eqref{haunn} the indices $\mu_1,\ldots,\mu_n$ are upper (contravariant)
Lorentz indices, $\mu_i=0,\ldots,d-1$. The index $a$ is distinguished from the latter. Of course  as long as the background metric is flat all indices are on the same footing, but later on $a$  will be interpreted as a tangent space index and $h_a$ will be called {\it frame-like master field}. It is therefore convenient to keep track of its special nature from the beginning.

The master field $h^a(x,u)$ is characterized by the following gauge transformation properties
\be
\delta_\varepsilon h_a(x,u) = \partial_a^x 
\varepsilon(x,u)-i [h_a(x,u) \stackrel{\ast}{,} \varepsilon(x,u)] 
\equiv {\cal D}^{\ast x}_a  \varepsilon(x,u),
\label{deltahxp}
\ee
where the covariant derivative
\be
{\cal D}^{\ast x}_a = \partial_a^x- i  [h_a(x,u) \stackrel{\ast}{,}\quad].\0
\ee 
has been introduced. The $\ast$ product is the Moyal product, defined as follows
\be
f(x,u)\ast g(x,u) = f(x,u) e^{\frac i2 \left( \stackrel{\leftarrow}{\partial}_x \stackrel{\rightarrow}{\partial}_u - \stackrel{\leftarrow}{\partial}_u\stackrel{\rightarrow}{\partial}_x\right)}g(x,u) \label{Moyal}
\ee
In the framework of the Weyl-Wigner quantization, the master field $h_a(x,u)$ would be identified as a symbol.

In order to emphasize the difference between the index $a$ and the others, it is useful to imitate the ordinary gauge theories using the compact notation  ${\bf d}= \partial_a\, dx^a, {\bf h}= h_a dx^a$ ($x^a$ are coordinates in the tangent spacetime) and  writing \eqref{deltahxp} as
\be
\delta_\varepsilon {\bf h}(x,u) ={\bf d} 
\varepsilon(x,u)-i [{\bf h} (x,u) \stackrel{\ast}{,} \varepsilon(x,u)] \equiv 
\bfD \varepsilon (x,u),\label{deltahxpbf}
\ee
where it is understood that 
\be
  [{\bf h} (x,u) \stackrel{\ast}{,} \varepsilon(x,u)]=  [h_a (x,u)
\stackrel{\ast}{,} \varepsilon(x,u)]dx^a,\0
\ee
Continuing on the same tune one can introduce the curvature notation
\be
{\bf G} = {\bf d} {\bf h}
 -\frac i 2 [ {\bf h}\stackrel {\ast}{,}{\bf h}],\label{curv1} 
\ee
which transforms as
\be
\delta_\varepsilon {\bf G} = -i [{\bf G}  \stackrel {\ast}{,}\varepsilon]
\label{deltaG}
\ee

Let us come next to the action for the master field $h_a(x,u)$. Like in ordinary gauge theories it will be an integrated polynomial of $\bf G$ or of its components $G_{ab}$. In order to exploit the transformation property \eqref{deltaG} like in ordinary gauge theories, we need the `trace property', analogous to the trace of polynomials of Lie algebra generators in ordinary non-Abelian gauge
theories.  In our context such an object is 
\be
\langle\!\langle f\ast g\rangle\!\rangle\equiv \int d^dx \int \frac
{d^du}{(2\pi)^d}
f(x,u)\ast g(x,u) = \int d^dx \int \frac {d^du}{(2\pi)^d} f(x,u) g(x,u)=
\langle\!\langle g\ast f\rangle\!\rangle  \label{trace}
\ee
From this, plus associativity, it follows that
\be
&&\langle\!\langle f_1 \ast f_2\ast \ldots \ast f_n\rangle\!\rangle=
\langle\!\langle f_1 \ast (f_2\ast \ldots \ast f_n)\rangle\!\rangle\0\\
&&=(-1)^{\epsilon_1(\epsilon_2+\ldots+\epsilon_n)} \langle\!\langle  (f_2\ast
\ldots \ast f_n)\ast
f_1\rangle\!\rangle=(-1)^{\epsilon_1(\epsilon_2+\ldots+\epsilon_n)} 
\langle\!\langle  f_2\ast \ldots \ast f_n\ast f_1\rangle\!\rangle\label{cycl}
\ee
where $\epsilon_i$ is the Grassmann degree of $f_i$. In particular
\be
\langle\!\langle [f_1 \stackrel{\ast}{,} f_2\ast \ldots \ast
f_n\}\rangle\!\rangle=0\label{comm0}
\ee
where $[\quad  \stackrel{\ast}{,}\quad\}$ is the $\ast$-commutator or
anti-commutator, as appropriate.

Incidentally, this property holds also when the $f_i$ are valued in a (finite dimensional) Lie algebra, provided the symbol $\langle\!\langle   \quad  \rangle\!\rangle$ includes also the trace over the
Lie algebra generators.

The curvature components, see \eqref{curv1}, are
\be
G_{ab}= \partial_a h_b - \partial_b h_a -i [h_a \stackrel{\ast}{,} h_b ]
\label{Gab}
\ee
with transformation rule
\be
\delta_\varepsilon G_{ab}=-i [G_{ab}\stackrel{\ast}{,}
\varepsilon]\label{deltaGab}
\ee
Therefore, the functional $\langle\!\langle G^{ab} \ast G_{ab} \rangle\!\rangle$ transforms like
\be 
\delta_\varepsilon \langle\!\langle G^{ab} \ast G_{ab} \rangle\!\rangle = -i
\langle\!\langle 
   G^{ab} \ast G_{ab} \ast\varepsilon -\varepsilon \ast  G^{ab} \ast G_{ab}
\rangle\!\rangle=0 \label{YMinvariance}
\ee
Therefore 
\be
{\cal YM}({\bfh})=- \frac 1{4 g^2}\langle\!\langle G^{ab} \ast G_{ab}
\rangle\!\rangle\label{YMh}
\ee
is invariant under HS gauge transformations and it is a well defined functional
in any dimension. .

\subsection{The non-Abelian case}

All that has been done so far for the Abelian case can be repeated for
the non-Abelian case with minor changes. One simply assumes that the master field $h_a$ belongs to the adjoint representation of a non-Abelian Lie algebra with generators $T^\alpha$. 
\be
{\mathsf h}= {\mathsf h}^\alpha T^\alpha, \quad\quad {\mathsf h}^\alpha =
h^\alpha_a
dx^a\label{halpha}
\ee
where summation over $\alpha$ is understood.  The HS gauge parameter is
\be
{\mathsf e} (x,u) = \varepsilon^\alpha(x,u) T^\alpha\label{epsilonnA}
\ee
and the transformation of $ {\mathsf h} (x,u) $  
\be
\delta_{\mathsf e} {\mathsf h} (x,u) = \bfd^x 
	{\mathsf e}(x,u)-i [{\mathsf h}(x,u) \stackrel{\ast}{,}{\mathsf e}
(x,u)] ,
\label{deltahxpnA}
\ee
assuming the generators $T^\alpha$ are anti-hermitean. Next the curvature is
\be
{\mathsf G} = {\bf d} {\mathsf h}
 -\frac i 2 [ {\mathsf h}\stackrel {\ast}{,}{\mathsf h}]\label{curv2}
\ee
where the $\ast$-commutator includes now also the Lie algebra commutator. Of course we
have, in particular,
\be
\delta_{\mathsf e} {\mathsf G} (x,u)= -i [{\mathsf G}(x,u)
\stackrel{\ast}{,}{\mathsf e} (x,u)]
\ee
Everything works as before provided, as noted above, the symbol $\langle\! \langle \quad
\rangle\!\rangle$ includes also the trace over the Lie algebra generators.
Finally
\be
{\cal YM}({\mathsf h})=- \frac 1{4 g^2}\langle\!\langle {\mathsf G}^{ab} \ast
{\mathsf G}_{ab}
\rangle\!\rangle\label{YMhnonAb}
\ee
is invariant under the HS non-Abelian gauge transformations and it is a well defined 
functional in any dimension.

The functionals \eqref{YMh} { and \eqref{YMhnonAb} differ from the usual field theory actions in that   they are integrals over the phase space of a point  particle with coordinate $x^a$. In particular  their  dimension is not the one of an ordinary field theory action, due to the integration over $u$. Thus, we should divide it by a factor ${\cal V}_u$  proportional to the integration volume over the momentum space. Since it does not play any role at the classical level, for the sake of simplicity, we understand this factor. 

Notwithstanding this difference we can extract from \eqref{YMh} covariant
eom's by  taking the variation with respect to $\bfh$. As was shown in \cite{HSYM}, 
we can assume that the action principle holds for fields defined in the phase space. 
From\eqref{YMh}  we get the following eom:
\be
\partial_b G^{ab} -i [h_b\stackrel{\ast}{,} G^{ab}]\equiv{\cal D}_b^\ast
G^{ab}=0\label{YMeom}
\ee
which is, by construction, covariant under the HS gauge transformation
\be
\delta_\varepsilon\left( {\cal D}_b^\ast G^{ab}\right)= -i[ {\cal D}_b^\ast
G^{ab},\varepsilon]\label{covYMeom}
\ee
The expressions of these eom's in components can be found \cite{HSYM}. For each component field $\varphi_a^{\mu_1\ldots\mu_s}$ they  take the form $0=\square \varphi_a^{\mu_1\ldots\mu_s}+\ldots$, with $s=0,1,2,...$. Ellipses  refer to terms containing two or more
derivatives and at most cubic in the component fields; at each order, defined by the number of
derivatives, there is a finite number of terms.  A  theory with this characteristic is called {\it
perturbatively local}
}.  

\subsection{Gauge transformations}

To understand the meaning of the component fields (at least, the lowest lying ones) in $h_a$ we examine 
in detail the gauge transformation \eqref{deltahxp}. Let us expand the master gauge parameter $\varepsilon(x,u)$
\be
\varepsilon(x,u)&=& \epsilon(x) +\xi^\mu(x) u_\mu+\frac 12
\Lambda^{\mu\nu}(x)u_\mu
u_\nu+\ldots\label{epsxu}
\ee
and let us consider the first few terms in the transformation of the lowest spin fields. We order them so that the component fields and gauge parameters are infinitesimals of the same order. For instance to lowest order the transformation \eqref{deltahxp} reads
\be
 \delta^{(0)} A_a= \partial_a \epsilon, \quad\quad 
\delta^{(0)} \chi_a^{\nu} = \partial_a \xi^\nu , \quad\quad \delta^{(0)}b_a{}^{\nu\lambda} = \partial_a\Lambda^{\nu\lambda}, \quad\quad \ldots\label{deltaAhb}
\ee
to first order we have
\be
\delta^{(1)} A_a &=& \xi\!\cdot\!\partial A_a - \partial_\rho \epsilon
\,\chi_a^{\rho} \label{delta1Ahb}\\
\delta^{(1)} \chi_a^{\nu} &=& \xi\!\cdot\!\partial \chi_a^\nu-\partial_\rho
\xi^\nu \chi_a^\rho 
+ \partial^\rho A_a \Lambda_{\rho}{}^\nu   - \partial_\lambda \epsilon
\,b_a{}^{\lambda\nu}\0\\
\delta^{(1)} b_a^{\nu\lambda} &=& \xi\!\cdot\!\partial b_a{}^{\nu\lambda}
-\partial_\rho \xi^\nu b_a{}^{\rho\lambda}- \partial_\rho \xi^\lambda 
b_a{}^{\rho\nu }+{\partial_\rho \chi_a^{\nu} \Lambda^{\rho\lambda}
+\partial_\rho
\chi_a^{\lambda} \Lambda^{\rho\nu}}
- \chi_a^{\rho} \partial_\rho { \Lambda^{\nu\lambda}} \0\\
\ldots&=&\ldots\0
\ee
The next orders contain three and higher derivatives.

These transformation properties suggest to associate the first two component fields of $h_a$ to an ordinary U(1) gauge field and to the fluctuating part of a vielbein. To see this
let us denote by $\tilde A_a, \tilde e_a^\mu = \delta_a^\mu -\tilde \chi_a^\mu$
the standard gauge and vielbein fields. The standard gauge and diff transformations, are
 \be
\delta \tilde A_a&\equiv& \delta \left(\tilde e_a^\mu \tilde A_\mu\right)\equiv
\delta  \left((\delta_a^\mu -\tilde \chi_a^\mu) \tilde
A_\mu\right)\label{standardtransf}\\
&=&\left(-\partial_a \xi^\mu -\xi\!\cdot\!\partial \tilde \chi_a^\mu +\partial_\lambda \xi^\mu
\tilde \chi_a^\lambda\right) \tilde A_\mu+(\delta_a^\mu -\tilde
\chi_a^\mu)\left(\partial_\mu\epsilon + \xi\!\cdot\! \partial \tilde A_\mu + \tilde A_\lambda \partial_\mu \xi^\lambda\right) \0\\
&= &
\partial_a\epsilon + \xi\!\cdot\! \partial \tilde A_a- \tilde \chi_a^\mu
\partial_\mu\epsilon \0 
\ee
and
\be
\delta \tilde e_a^\mu \equiv  \delta  (\delta_a^\mu -\tilde \chi_a^\mu) =
\xi\!\cdot\! \partial\tilde e_a^\mu -\partial_\lambda \xi^\mu \tilde e_a^\lambda = -
\xi\!\cdot\!\partial \tilde \chi_a^\mu -\partial_a \xi^\mu +\partial_\lambda \xi^\mu
\tilde \chi_a^\lambda\label{deltaeamu}
\ee
so that
\be
\delta \tilde \chi_a^\mu= \xi\!\cdot\! \partial\tilde \chi_a^\mu +\partial_a \xi^\mu
-\partial_\lambda \xi^\mu \tilde \chi_a^\lambda\label{deltaeamu1}
\ee
where we have retained only the terms at most linear in the fields.

Now it is crucial to interpret the derivative $\partial_a$ in eq.\eqref{deltahxp} and \eqref{deltaAhb} in the appropriate way:  the
derivative $\partial_a$ means $\partial_a = \delta_a^\mu \partial_\mu,$ not
{ $ \partial_a = e_a^\mu \partial_\mu= \left(\delta_a^\mu
-\chi_a^\mu\right)\partial_\mu$.}
for the linear correction $ -\chi_a^\mu\partial_\mu$ is
contained in the term $ -i [h_a(x,u) \stackrel{\ast}{,} \varepsilon(x,u)]$, see
for instance the second term in the RHS of the first equation
\eqref{delta1Ahb}. From this it is straightforward to make the identifications
\be
A_a= \tilde A_a, \quad\quad \chi_a^\mu = \tilde \chi_a^\mu\label{identAAee}
\ee

The transformations \eqref{deltaAhb}, \eqref{delta1Ahb}
allow us to interpret  $\chi_a^\mu$ as the fluctuation of the inverse 
vielbein. Although the name master frame-like field for $h_a(x,u)$ is somewhat of an abuse of language (because in fact $h_a$ contains only the fluctuating part of the frame field) we shall refer to it with this name.

\section{Local Lorentz symmetry}
\label{s:LLI}

From the above identification we deduce that the HS-YM-like action may accommodate gravity. 
However a gravitational interpretation requires the frame field  to transform also under local Lorentz transformations. Therefore one would expect the master field $h_a$ to transform and the action to be invariant under local Lorentz transformations { (LLT)}. This is apparently not so for the action \eqref{YMh},  and, at least at first sight, the local Lorentz invariance does not seem to be there.

Let us consider the simple case in
which only the field $A_a$ is non-vanishing, {so that} the form of the Lagrangian is
\be
L_A \sim F_{ab} F^{ab}, \quad\quad F_{ab} =\partial_a A_b-\partial_b
A_a\label{LA}
\ee
This is not invariant under a local Lorentz transformation, because, under $A_a \to 
A_a+\Lambda_a{}^b A_b$, { the} terms $\left(\left(\partial_a
\Lambda_b{}^c\right) A_c -\left( \partial_b \Lambda_a{}^c
\right)A_c\right)F^{ab}$ are generated, that do not vanish. In the sequel we would like to show that
the local Lorentz symmetry is present in our formulation of HS-YM like theories,  albeit in a hidden form. It is however necessary to change perspective and to resort to a different formulation of the geometry of gravity, called {\it teleparallelism}.

Let us start {   by defining a trivial (inverse) frame
$e_a^\mu (x)$ as} a frame that can be reduced to a Kronecker delta by means of a
local Lorentz transformation, i.e.  
such that there exists a (pseudo)orthogonal transformation $O_a{}^b(x)$ for
which
\be
O_a{}^b(x) e_b{}^\mu (x) = \delta_a^\mu\label{trivialframe}
\ee
As a consequence $e_b{}^\mu (x)$ contains only inertial (non-dynamical)
information. A full gravitational (dynamical) frame is the sum of a trivial frame and a
nontrivial piece
\be
\hat E_a^\mu(x) = e_a{}^\mu (x)-\hat\chi_a^\mu(x)\,. \label{fullframe}
\ee
By means of the above LLT it can be reduced to the form
\be
E_a^\mu(x) = \delta_a^\mu -\chi_a^\mu(x)\,, \label{fullframetr}
\ee
{where $\chi_a^\mu(x)=O_a{}^b(x)\hat\chi_b^\mu(x)$.}
This is the form we have encountered above in HS theories. In this framework
the Kronecker delta represents a trivial frame. If we want to
recover local Lorentz covariance, instead of $\partial_a=\delta_a^\mu \partial_\mu$ we must understand 
\be
\partial_a = e_a{}^\mu(x) \partial_\mu, \label{truepartiala}
\ee
where $e_a{}^\mu(x)$ is a trivial (or purely inertial) vielbein. In particular,
under an infinitesimal LLT, it transforms according to
\be
\delta_\Lambda  e_a{}^\mu(x) = \Lambda_a{}^b(x) e_b{}^\mu(x)\label{deltaLea}
\ee 

In a similar way we can define a trivial connection (or inertial spin connection) by
\be
\EA^a{}_{b\mu} = \left(O^{-1}(x) \partial_\mu O(x)\right)^a{}_b\label{teleA}
\ee
where $O(x)$ is a generic local (pseudo)orthogonal transformation (finite local 
Lorentz transformation).  The corresponding curvature vanishes
\be
\ER^a{}_{b\mu\nu} = \partial_\mu \EA^a{}_{b\nu}- \partial_\nu \EA^a{}_{b\mu} 
+ \EA^a{}_{c\mu}\EA^c{}_{b\nu}- \EA^a{}_{c\nu}\EA^c{}_{b\mu}=0\label{teleER}
 \ee
Now, the space of connection is affine, i.e. we can obtain any
connection from a fixed one by adding to it
tensors that transform according to the adjoint representation. When the
spacetime is topologically trivial we can choose as origin of the affine space
the 0 connection. The latter is a particular member in the class of trivial
connections as can be easily verified. Let us start with the  connection 
\eqref{teleA},  and act on it with a finite Lorentz transformation   $L(x)$ we get
\be
\EA_\mu(x) \rightarrow L(x) D_\mu L^{-1} (x)= L(x) (\partial_\mu + \EA_\mu)
L^{-1} (x)\label{LLL}
\ee
If we choose $L=O$ we get
\be
\EA_\mu(x) \rightarrow 0\label{Lfixing}
\ee
After this operation the {local Lorentz} symmetry has disappeared, i.e. choosing the zero
spin connection amounts to fixing the local Lorenz gauge.

$\EA_\mu$ is nevertheless a connection and it makes sense to define the inertial
covariant derivative
\be
D_\mu =  \partial_\mu +\EA_\mu\label{inertialcovder}
\ee
which is Lorentz covariant. In ordinary Riemannian geometry the vielbein is annihilated by the covariant
derivative provided we use it to build the metric  and consequently the
Christoffel symbols.
A trivial frame and a trivial connection have an analogous relation provided the
(pseudo)orthogonal transformation $O$ in \eqref{trivialframe} and \eqref{teleA}
is the same in both cases. For we have
\be
D_\mu e_a^\nu&= &\left(\partial_\mu \delta_a^b+ \EA_{\mu a}{}^b \right) e_b^\nu
=
\partial_\mu e_a^\nu +  \left( O^{-1} \partial_\mu O\right) _a{}^b O_b^{-1}{}^c
\delta_c^\nu\0\\
&=& \partial_\mu O_a^{-1}{}^c \delta_c^\nu - \partial_\mu O_a^{-1}{}^c
\delta_c^\nu =0
\label{metriclike}
\ee
From now on we assume that this is the case.

The connection $\EA_\mu$ contains only inertial and no gravitational
information. It will be referred to as the {\it inertial connection}. It is a
{\it non-dynamical} object (its content is pure gauge), and plays a role
analogous to a trivial frame $e_a^\mu(x)$, considered before. The dynamical degrees of freedom will be contained in the
adjoint tensor to be added to $\EA_\mu$ in order to form a fully dynamical spin 
connection. The splitting, illustrated above,  of vierbein and spin connection into an
inertial  and a dynamical part goes under the name of teleparallelism, see
\cite{teleparallel}.

To complete our LL covariantization process of the HS-YM like theories, let us add that replacing simple spacetime derivatives $\partial_\mu$ with the inertial ones
$D_\mu$ everywhere is not enough. There is also another apparent inconsistency. Let us take, as an example, the HS field strength $G_{ab}$, \eqref{Gab}.
Following the above recipe we must replace everywhere, also in the $\ast$
product, the ordinary derivatives with covariant ones (covariant with respect to
the spin connection $\EA_a$)\footnote{Replacing $\partial_\mu$ with $D_\mu$ 
does not create any ordering problem because $[D_\mu,D_\nu]=0$.}. This changes the transformation properties for the various pieces. For instance $  D_a h_b $ transforms differently from 
\be
\delta_\Lambda (  h_a \ast h_b) = \Lambda_a{}^c (h_c \ast h_b) + \Lambda_b{}^c 
h_a \ast h_c
\label{deltahahb}
\ee
The inertial frame fixes this inconsistency. Instead of writing
$\partial_a=\delta_a^\mu \partial_\mu$ we must write $\partial_a =
e_a{}^\mu(x) \partial_\mu$, 
where $e_a{}^\mu(x)$ is a purely inertial frame. In particular, under a LLT, it
transforms according to $ \delta_\Lambda  e_a{}^\mu(x) = \Lambda_a{}^b e_b{}^\mu(x)$.

Similarly, whenever a flat index $O_a$ is met, we must rewrite it $O_a
=e_a{}^\mu  O_\mu $. Finally in spacetime integrated expression we must introduce in the integrand the factor $e^{-1}$, where $e= \det \left(e_a^\mu\right)$, the determinant of
the inertial frame.

With such recipes all inconsistencies disappear. For instance
\be
\delta_\Lambda (D_a J_b)= \Lambda_a{}^c  (D_c J_b)+ \Lambda_b{}^c  (D_a J_c)\0
\ee
Therefore $\delta_\Lambda(\eta^{ab}D_a J_b)=0$.  

Likewise 
\be
\delta_\Lambda G_{ab} = \Lambda_a{}^c G_{cb} +  \Lambda_b{}^c
G_{ac}\label{deltaLGab}
\ee
which implies the local Lorentz invariance of $G_{ab} G^{ab}$ and of the full action \eqref{YMh}.

It should be emphasized that this is a new approach to gravity. The HS gauge
transformation \eqref{deltahxp} reproduces both ordinary U(1) gauge transformations and
diffeomorphisms, but the eom of $\chi_a^\mu$ are not quite the ordinary
metric equations of motion: the linear eom coincides with the ordinary one after gauge
fixing, but there is a huge difference with ordinary gravity because in the latter the
interaction terms are infinite and include all powers in the fluctuating field, while in the
\eqref{YMh} there are at most cubic interaction terms.

\section{Scalar and spinor master fields and BRST quantization}

We can couple to the HS YM-like theories matter-type fields of any spin, for instance, 
a complex multi-boson field 
\be
\Phi(x,u)= \sum_{n=0}^\infty \frac 1{n!}\Phi^{\mu_1\mu_2\ldots \mu_n}(x)
 u_{\mu_1}u_{\mu_2}\ldots u_{\mu_n}\label{Phixu}
\ee
which, under a master gauge transformation \eqref{deltahxp} transforms like $
\delta_\varepsilon \Phi = i \varepsilon\ast\Phi$.
We define as well the covariant derivative $ \ED^\ast_a \Phi= \partial_a \Phi -i h_a \ast \Phi$, which has the property $\delta_\varepsilon \ED^\ast_a \Phi= i\,\varepsilon \ast \ED^\ast_a \Phi$.
It follows that the kinetic action term $
\frac 12 \langle\!\langle(\ED_\ast^a
\Phi)^\dagger\ast\ED^\ast_a \Phi\rangle\!\rangle$ and potential terms 
$ \langle\!\langle(\Phi^\dagger \ast \Phi)^n_\ast\rangle\!\rangle$ are
 gauge invariant.

\vskip 1cm

In a very similar manner we can introduce master spinor fields,
\be
\Psi(x,u) = \sum_{n=0}^\infty\frac 1{n!} \Psi_{(n)}^{\mu_1\ldots \mu_n}(x)
u_{\mu_1} \ldots u_{\mu_n}, \label{Psi}
\ee
where $\Psi_{(0)}$ is a Dirac field. Under HS gauge transformations it transforms according to
$\delta_\varepsilon \Psi = i \varepsilon\ast\Psi $, so we can define the covariant derivative
\be
\ED^\ast_a \Psi= \partial_a \Psi -i h_a \ast \Psi\label{covderPsi}
\ee
so that $\delta_\varepsilon (\ED^\ast_a \Psi)=i\varepsilon \ast (\ED^\ast_a
\Psi)$.
It is evident that the action
\be
S(\Psi,h) = \langle\!\langle \overline \Psi i\gamma^a \ED_a \Psi
\rangle\!\rangle
=  \langle\!\langle \overline \Psi \gamma^a\left(i\partial_a+ h_a \ast\right)
\Psi \rangle\!\rangle
\label{Spsih}
\ee
is invariant under the HS gauge transformations \eqref{deltahxp}. 

By adding a scalar field sector with suitable potential it is possible to implement the Higgs mechanism, see \cite{HSYM}.
\vskip 1cm
This completes the presentation of the classical aspects of HS YM models. The rest of the paper is devoted to quantization. We start with the BRST formulation  of the models and then we present the perturbative quantization.

\subsection{BRST quantization of HS Yang-Mills}
\label{ssec:BRST}

To quantize the action \eqref{YMh} we must fix the gauge and apply the
Faddeev-Popov approach. We impose the Lorenz gauge with parameter $\alpha$ and apply the usual gauge theory approach. The quantum action becomes
\be
{\cal Y}{\cal M}(h_a,c,B)=\frac 1{g^2} \langle\!\langle  -\frac 1{4 } G_{ab}
\ast G^{ab} -
h^a\ast \partial_a B-i \partial^a \overline c\ast {\cal D}_a^\ast c +\frac
{\alpha}2 B\ast B  \rangle\!\rangle\label{YMhquantum}
\ee
where $c, \overline c$ and $B$ are the ghost, antighost and Nakanishi-Lautrup
master fields, respectively. $c, \overline c$ are anticommuting fields, while
$B$ is commuting.

The action \eqref{YMhquantum} is invariant under the BRST transformations
\be
s h_a ={\cal D}_a^\ast c,\quad\quad s c= i c\ast c = \frac i2 [c \stackrel{\ast}{,} c], \quad\quad
s \overline c  = i B, \quad\quad s B=0\label{BRSTtr}
\ee
which are nilpotent. It follows in particular
\be
s( {\cal D}_a^\ast c)=0, \quad\quad s (c\ast c)=0\0
\ee

From the point of view of the $u$ dependence $c, \overline c$ and $B$ are to be
expanded as in eq.\eqref{Phixu}.  Integrating out $B$ in \eqref{YMhquantum} we obtain the standard gauge-fixed form of the action. 
\be
{\cal Y}{\cal M}(h_a,c)=\frac 1{g^2} \langle\!\langle  -\frac 1{4 } G_{ab} \ast
G^{ab} -
\frac 1{2\alpha} \partial_a h^a\ast\partial_b h^b -i \partial^a \overline c\ast
{\cal D}_a^\ast c  \rangle\!\rangle\label{YMhquantum'}
\ee

\section{Perturbative quantization}

At this point we would like to produce formulas in order to be able to compute physical amplitudes following the scheme of perturbative quantization in ordinary gauge theories. This requires finding the propagators of the various component fields and identifying the vertices. To do so we have to come to terms with the $u$ integration. We proceed as follows. We first rescale fields, coordinates and the coupling constant $g$, so as  to factor out an innocuous integral over $u$.   Let us start by splitting the action \eqref{YMhquantum} in quadratic, cubic and quartic parts $ S_2+S_3+S_4$.
Then we split $\sqrt{u^2}= \mathfrak{m} {\rm u} $ where
$ {\mathfrak m}$ is a fixed mass scale and 
$\rm u$ is the dimensionless variable part, and we redefine $h_a^{\mu_1\ldots\mu_{s-1}}\to
h_a^{'\mu_1\ldots\mu_{s-1}}= {\rm u}^{s-1} h_a^{\mu_1\ldots\mu_{s-1}}$, 
the coordinates $ x^\mu \to {\rm u}  x^\mu$, and the coupling $g \to \frac g{\rm u}$. Under these redefinition the kinetic term, written as $ \langle\!\langle h_a K^{ab} h_b  \rangle\!\rangle$,  
remains the same apart from 
\be
\langle\!\langle h_a K^{ab} h_b  \rangle\!\rangle \longrightarrow
\langle\!\langle h_a K^{ab} h_b  \rangle\!\rangle' \label{kineticfinal}
\ee
with $u$ replaced by ${\mathfrak m}$. 
The symbol $  \langle\!\langle\quad \rangle\!\rangle'$ means that the
integration measure has changed to
\be
\int d^dx d^d u \longrightarrow
{\mathfrak m}^d\int d^dx d^d{\rm u} \, {\rm u}^{d-2}\label{intmeasure1}
\ee
Also $S_3$ and $S_4$ preserve the same form with $u$ replaced by
${\mathfrak m}$ and $\langle\!\langle \quad \rangle\!\rangle$ replaced by $
\langle\!\langle \quad \rangle\!\rangle'$. 

In other words, apart from this change of measure and the substitution of $u$
replaced by
${\mathfrak m}$, in the expressions $S_3,S_4$ and the kinetic term, nothing has
changed. In particular the dependence on ${\rm u}$ has disappeared from the
integrand.
Since now the integrand is  ${\rm u}$ independent we can factor out the quantity
\be
{\cal V}_d =   {\mathfrak m}^d\int d^d{\rm u} \, {\rm
u}^{d-2}\label{intmeasure2}
\ee
and simply get rid of it.

We  refer to this configuration of the theory, in which a mass parameter ${\mathfrak m}$ is evoked, the {\it frozen momentum frame}. It is our framework for quantization.

Finally we are
simply left with the spacetime action ${\cal S}= {\cal S}_2+  {\cal S}_3+ {\cal
S}_4$:
\be
{\cal S}_2= \int d^dx \sum_{\{\mu\},\{\nu\}} h^{Ta}_{ \{\mu\}}(x)
K_{ab}^{\{\mu\}\{\nu\}}(x,{\mathfrak m})  h^b_{ \{\nu\}}(x)\label{kineticnew} 
\ee
where $h^{Ta}_{\{\mu\}} = (A^a, \chi^a_\mu, b^a_{\mu_1\mu_2},
c^a_{\mu_1\mu_2\mu_3},\ldots)$. 
\be
{\cal S}_3&=&  {-g}\int d^dx\, \biggl{\{} \partial^a A^b (\partial_\sigma A_a
\chi_b^\sigma - \partial_\sigma A_b \chi_a^\sigma)\label{S3final}\\
&&-\frac 1{{24}} (\partial^a A^b-\partial^b A^a) \bigl(\partial_{\sigma_1}
\partial_{\sigma_2} \partial_{\sigma_3}A_a \, c_b^{\sigma_1\sigma_2\sigma_3}+ 3
\partial_{\sigma_3} b_a^{\sigma_1\sigma_2}\partial_{\sigma_1}
\partial_{\sigma_2}\chi_b^{\sigma_3}\bigr)\0\\
&&+ \frac {{\mathfrak m}^2}{2d} \Big(\partial^a A^b \partial_\sigma
b_{a\mu}{}^\mu
\chi_b^\sigma -\partial^a A^b \partial_\sigma b_{b\mu}{}^\mu \chi_a^\sigma+ \dots\Big)+\ldots\biggr{\}}\0
\ee
 and
\be
{\cal S}_4
&=& -  \frac {g^2}{{2}}\, \int d^dx \biggl{\{}  \bigl(\partial_\sigma A^a
\chi^{b\sigma}- \partial_\sigma A^b \chi^{a\sigma}\bigr) \partial_\tau A_a
\chi_b^\tau\label{S4final}\\
&& + \frac {{\mathfrak m}^2}{d} \Big(\bigl(\partial_\sigma A^a \chi^{b\sigma}-
\partial_\sigma A^b \chi^{a\sigma}\bigr) \bigl(\partial_\tau A_a
c_{b\nu}{}^{\tau\nu}+2\partial_\tau \chi_a^\nu b_{b\nu}{}^\tau +  \partial_\tau
b_{a\nu}{}^\nu\chi_b^\tau\bigr)+\ldots\biggr{\}}\0
\ee 

Now, the kinetic operator in \eqref{kineticnew} is
\be
&&K_{ab}^{\{\mu\}\{\nu\}}(x,{\mathfrak m})=\left( \eta_{ab} \square_x - \frac
{\alpha-1}\alpha
\partial^x_a\partial^x_b\right)N^{\{\mu\}\{\nu\}}({\mathfrak m})\equiv
\EK^x_{ab} N^{\{\mu\}\{\nu\}}({\mathfrak m})\label{Kabx1}
\ee
with 
{\scriptsize
\be
&& N^{\{\mu\}\{\nu\}}({\mathfrak m})=\label{Nm}\\
&&\left(\begin{matrix}
       1 & 0 &\eta^{\nu_1\nu_2} \frac {{\mathfrak m}^2}{2d} &0&
\Pi^{\nu_1\nu_2\nu_3\nu_4} \frac
{{\mathfrak m}^4}{4!d(d+2)}&0\\
0 &\eta^{\mu_1\nu_1} \frac {{\mathfrak m}^2}{d}& 0 &\Pi^{\mu_1\nu_1\nu_2\nu_3}
\frac
{{\mathfrak m}^4}{3!d(d+2)} &0&\ldots\\
\eta^{\mu_1\mu_2} \frac {{\mathfrak m}^2}{2d}& 0& \Pi^{\mu_1\mu_2\nu_1\nu_2}
\frac
{{\mathfrak m}^4}{4d(d+2)}&0&\ldots&\ldots\\
0 &\Pi^{\mu_1\mu_2\mu_3\nu_1} \frac {{\mathfrak m}^4}{3!d(d+2)} &0
&\ldots&\ldots&\ldots\\
 \Pi^{\mu_1\mu_2\mu_3\mu_4} \frac {{\mathfrak m}^4}{4!d(d+2)}&0&\dots
&\ldots&\ldots&\ldots\\
0 &\dots &\ldots&\ldots&\ldots&\ldots\\
 \end{matrix}\right)\0
\ee }
where $ \Pi_{\mu\nu\lambda\rho}= \eta_{\mu\nu}\eta_{\lambda\rho} + 
\eta_{\mu\lambda}\eta_{\nu\rho}+ \eta_{\mu\rho} \eta_{\nu \lambda}$.
{If the inverse of this matrix exists} the propagator in momentum space is 
 \be
&& \widetilde P^{\{\mu\}\{\nu\}}_{ab}(k,{\mathfrak m}) =-i
 \left(\frac {\eta_{ab}}{k^2} 
+(\alpha-1) \frac {k_ak_b}{k^4} \right)
\,M^{\{\mu\}\{\nu\}}({\mathfrak m})\label{prophahbtilde}
\ee
where $M$ is the inverse of $N$, i.e.
\be
N^{\{\mu\}\{\nu\}}({\mathfrak m}) M_{\{\nu\}\{\lambda\}}({\mathfrak m})=
\delta^{\{\mu\}}_{\{\lambda\}}\0
\ee
One can rapidly verify that the matrix $N$ is not invertible. To achieve invertibility we must restrict ourselves to traceless component fields, in which case both matrices $N$ and $M$ are diagonal and
\be
\widetilde M^{\{\mu\}}_{\{\nu\}}({\mathfrak m})=
\left(\begin{matrix}
       1 & 0 & 0&0&0&\ldots\\
0 &\delta^{\mu_1}_{\nu_1}\frac {1}{{\mathfrak m}^2}& 0 &0 &0&\ldots\\
0& 0& \Delta^{ \mu_1\mu_2}_{\nu_1\nu_2}
\frac{1}
{{\mathfrak m}^4}&0&0&\ldots\\
0 &0 &0 &\Delta^{ \mu_1\mu_2\mu_3}_{\nu_1\nu_2\nu_3}
\frac{1}
{{\mathfrak m}^6}&0&\ldots\\
0 &0 &0 &0&\Delta^{  \mu_1\mu_2\mu_3\mu_4}_{\nu_1\nu_2\nu_3\nu_4}
\frac{1}
{{\mathfrak m}^8}&\ldots\\
\ldots&\dots &\ldots&\ldots&\ldots&\ldots\\
 \end{matrix}\right)\label{tildeMm}
\ee
where 
\be
\Delta^{(n)\mu_1\ldots \mu_n}_{\quad\nu_1\ldots \nu_n}\equiv \Delta^{\mu_1\ldots \mu_n}_{\nu_1\ldots \nu_n}=\frac 1{n!}\left( 
\delta^{\mu_1}_{\nu_1}\ldots
\delta^{\mu_n}_{\nu_n}+{\rm perm} (\mu_1,\ldots,\mu_n)\right)\label{Delta}
\ee
The tilde on the matrix $M$ is to remember the restriction to a traceless basis. As will appear soon this restriction is not arbitrary, it indeed necessary in order to guarantee absence of ghosts.

\section{Physical and unphysical modes, hidden and manifest symmetry}

The HS-YM models abound of unphysical modes. Some of them, related to the index $a$ of $h_a$ , are taken care of by the BRST quantization and canceled by the FP ghosts. But there are many others related to the indices $\mu_1,\ldots,\mu_n$, which are visible in the form of the propagator \eqref{tildeMm}. To be precise by physical modes of 
$h_a^{\mu_1\ldots\mu_n}$ we understand solutions of its free massless equation of motion represented by a plane wave multiplied by a polarization, say $\theta_a^{\mu_1\ldots\mu_n}$, with 
non-negative norm: $\theta_a^{\mu_1\ldots\mu_n} \theta^a_{\mu_1\ldots\mu_n}\geq 
0$. To distinguish these {\it physical modes} from the strictly transverse ones,  i.e. $h_i^{j_1\ldots j_n}$,  we call the latter {\it strictly physical}.

How do we get rid of these unphysical modes and how do we guarantee Lorentz covariance? 
The argument is simple. It consists in decomposing the tensor 
$\Delta^{(n)}_{\mu_1\ldots \mu_n,\nu_1\ldots \nu_n}$, into a sum of orthogonal 
projectors, each corresponding to a representation of  the 
Lorentz group. Only one projector in this sum projects to physical states, 
which are traceless and transverse. Therefore replacing in \eqref{tildeMm}  the 
identity  $\Delta^{(n)}_{\mu_1\ldots \mu_n,\nu_1\ldots \nu_n}$ with this unique 
(Lorentz covariant) projector guarantees that, in amplitudes with physical 
states in the external legs, only physical states propagate in the internal line, thus ensuring the 
absence of propagating unphysical modes in any physical process.

Let us introduce the elementary projectors
\be
\pi_{\mu\nu}= \eta_{\mu\nu} - \frac {k_\mu k_\nu}{k^2} , \quad 
\quad\omega_{\mu\nu} =\frac {k_\mu k_\nu}{k^2} \label{elemproj}
\ee
with the properties
\be
\pi_{\mu\nu}\, \pi^{\nu}{}_\lambda= \pi_{\mu\lambda},\quad\quad 
\omega_{\mu\nu}\, \omega^{\nu}{}_\lambda= \omega_{\mu\lambda},\quad\quad 
\pi_{\mu\nu}\, \omega^{\nu}{}_\lambda=0  \label{elemproj1}
\ee
$\pi$ is transverse, while $\omega$ is not.

Let us see, as an example, the well-known case $n=2$:
\be
\Delta^{(2)}_{\mu_1\mu_2,\nu_1\nu_2} = P^{(2)}_{\mu_1\mu_2,\nu_1\nu_2}+ 
P^{(1)}_{\mu_1\mu_2,\nu_1\nu_2}+ P^{(0)}_{\mu_1\mu_2,\nu_1\nu_2}+ \overline 
P^{(0)}_{\mu_1\mu_2,\nu_1\nu_2}\label{spin2}
\ee
where
\be
P^{(2)}_{\mu_1\mu_2,\nu_1\nu_2}&=& \frac 12 
\left(\pi_{\mu_1\nu_1}\pi_{\mu_2\nu_2} +\pi_{\mu_1\nu_2}\pi_{\mu_2\nu_1} 
\right)-\frac 1{d-1} \pi_{\mu_1\mu_2}\pi_{\nu_1\nu_2}\label{spin2a}\\
P^{(1)}_{\mu_1\mu_2,\nu_1\nu_2}&=& \frac 12 
\left(\pi_{\mu_1\nu_1}\omega_{\mu_2\nu_2} +\pi_{\mu_1\nu_2}\omega_{\mu_2\nu_1}+ 
\omega_{\mu_1\nu_1}\pi_{\mu_2\nu_2} +\omega_{\mu_1\nu_2}\pi_{\mu_2\nu_1}\right) 
\label{spin2b}\\
P^{(0)}_{\mu_1\mu_2,\nu_1\nu_2}&=&\frac 1{d-1} 
\pi_{\mu_1\mu_2}\pi_{\nu_1\nu_2}\label{spin2c}\\
\overline P^{(0)}_{\mu_1\mu_2,\nu_1\nu_2}&=& 
\omega_{\mu_1\mu_2}\omega_{\nu_1\nu_2}\label{spin2d}
\ee
These are projectors orthogonal to one another. $P^{(2)}$ is transverse and traceless.  $P^{(1)}$ is 
traceless but not transverse,  $P^{(0)}$ is transverse but not traceless,  
$\overline P^{(0)}$ is neither transverse nor traceless. The relevant physical projector is of course $P^{(2)}$. This decomposition can be generalized. For any $n$ we can extract from $\Delta^{(n)}$ a unique transverse traceless projector ${\cal P}^{(n)}$, see \cite{HSYM}:
\be
 \Delta^{(n)}_{\mu_1\ldots \mu_n,\nu_1\ldots \nu_n}=\EP^{(n)}_{\mu_1\ldots 
\mu_n,\nu_1\ldots \nu_n}+\sum_{i=1}^{p_n} P^{(i)}_{\mu_1\ldots 
\mu_n,\nu_1\ldots 
\nu_n}
\label{PbarPtrtr}
\ee
where $\EP^{(n)}$ is transverse and traceless, while the remaining  $ P^{(i)}$ 
are traceful or non-transverse or both. The projector $ \EP^{(n)}$ is the only one 
that  projects onto the physical degrees of freedom. All the others project to 
nonphysical modes.All the above projectors are mutually orthogonal. It is worth stressing that 
they are all Lorentz covariant. 

Inserting in Feynman amplitudes only propagators where $\Delta^{(n)} $ are replaced by $ \EP^{(n)}$ guarantees that only the physical modes propagate (and no physical modes are excluded). So far we have considered only the $h_a$ component fields, but one can easily see that the same rules can be applied also to ghost and matter fields.

\subsection{Hidden and manifest gauge symmetry}

The prescription given above for constructing physical amplitudes, i.e. using Feynman diagram where
the ordinary propagators are replaced by new ones containing the transverse and traceless projectors  $ \EP^{(n)}$, guarantees that only physical modes propagate in the internal lines (while preserving Lorentz covariance). In other words, it guarantees unitarity for first order amplitudes. This raises two questions. The first is: we are familiar with the mechanism that ties the elimination of unphysical modes to the existence in the theory of a gauge symmetry, is it the same in HSYM theories and in what sense? Returning to the master field $h_a(x,u)$,
to compensate for the unphysical degrees of freedom in such theories we would need a symmetry under local transformations of the master field $h_a$ components, parametrized as follows
\be
\delta h_a^{\mu_1\ldots\mu_n}(x) \sim \partial^{(\mu_1} \Lambda_a^{\mu_2\ldots\mu_n)}(x)+\ldots, \quad\quad n\geq 2.
\label{newgaugetransf}
\ee
linear in $\Lambda_a^{\mu_2\ldots\mu_n}$, but with possible additional terms represented by the ellipses. The HS-YM action is clearly not invariant under such transformations. The point is that, similarly to local Lorentz invariance, this gauge invariance is completely fixed in the defining action \eqref{YMh}. But similar to local Lorentz symmetry, this additional gauge symmetry can be implemented by suitably modifying the original action: just replace everywhere in the action any component field $h_a^{\mu_1\ldots\mu_n}$ with $n\geq 2$ with its projection by the appropriate $\EP^{(n)}$ in configuration space. This means that $\pi_{\mu\nu}$ is replaced by 
\be
\check \pi_{\mu\nu}(\partial) = \eta_{\mu\nu} - \frac{\partial_\mu\partial_\nu}{\square} \label{tildepi}
\ee
For instance the third component field $b_a^{\mu\nu}$ becomes 
\be
\check b_a= b_a-\frac {\partial}{\square}\, \partial\!\cdot \!b_a + \frac {\partial^2}{\square^2}\partial\!\cdot\!\partial\!\cdot \! b_a -\frac 1{d-1} \left( \eta - \frac {\partial^2} {\square} \right) \left(b_a'-\frac 1{\square} \partial\! \cdot\!\partial\!\cdot\! b_a\right) \label{tildeba}
\ee
where we have adopted the compact notation of \cite{Francia:2002aa}, where the upper symmetric indices are understood, a dot, $\cdot$, denotes index contraction, a prime $'$ denotes a trace, free $\partial$ a gradient.

The same can be done for all the components of $h_a$ with more than 2 upper indices.
After such replacements \eqref{YMh} becomes a nonlocal action, let us call it ${\cal Y}{\cal M}(\check {\bf h})$, which is, however, automatically invariant under 
\be
\delta h_a^{\mu_1\ldots\mu_n}(x) = n\,\partial^{(\mu_1} \Lambda_a^{\mu_2\ldots\mu_n)}(x), \quad\quad n\geq 2.
\label{hiddengaugetransf}
\ee
This follows from the fact that all $\EP^{(n)}$ are transverse to the momentum and from the form of  \eqref{hiddengaugetransf} where a  derivative always factors out. The nonlocality can possibly be reabsorbed by adding suitable auxiliary fields, much like in \cite{Francia:2002aa} (for free HS theories).

But now comes the second question: a procedure like the one just outlined can in principle be applied to any theory; what makes HS-YM like theories different from a generic one? Things may go well at the lowest order, but what happens at the next orders? This question clearly calls forth renormalization (together with unitarity). 

We remark that the new action ${\cal Y}{\cal M}({\bf {\check h}})$ has the same form as \eqref{YMh}. Therefore it is invariant also under the new HS gauge transformations 
\be
\delta_{\hat\varepsilon} \check h_a(x,u)= \partial_a {\check \varepsilon}(x,u)     -i [\check h_a(x,u) \stackrel{\ast}{,} \check \varepsilon(x,u)] \label{newHSgaugetransf}
\ee   
So the new action is invariant both under these HS gauge transformations and under \eqref{hiddengaugetransf}.
The relation between $\check \varepsilon$ and the $\varepsilon$ of the original HS gauge transform \eqref{deltahxp} may be very complicated, and, in particular, field dependent and nonlocal. However what matters is that such a symmetry exists in the new action and allows us to write down the corresponding Ward identities, which are the building blocks of renormalization. This is our future challenge.

\section{Conclusions}

This paper is a progress report on massless HS-YM-like theories in flat spacetime.
We have shown that such theories can be constructed in any dimension, we have
defined their actions and found their equations of motion. They are perturbatively local. They are
characterized by a HS gauge symmetry which includes in particular ordinary gauge
transformations and diffeomorphisms.  Although initially the local Lorentz symmetry is missing, we have shown how to recover it, so that it is legitimate to say that these theories can incorporate gravity. Much like in ordinary gauge theories, with the addition of ghosts and auxiliary fields, they can be easily BRST quantized. Also the Higgs mechanism can be reproduced. HS-YM-like theories can also be presented in a supersymmetric version, in which rigid supersymmetry coexists with gravity, see \cite{bonora2020}.

We have also taken on the problem of the perturbative quantization of HS YM-like
models. We have seen that a 
perturbative expansion and the relevant Feynman diagrams can be defined.  
In this context a crucial issue is represented by the unphysical modes. In a 
gauge theory they are unavoidable, and may lead to unitarity violations; but good theories 
contain the necessary antidotes. This seems to be the case also for the 
HS YM-like theories. We have shown that the quantum perturbative series for 
physical amplitudes can be formulated in such a way as to exclude unphysical 
modes and allow only the propagation of the physical ones. This remarkable
result has been obtained by using both the traditional FP ghosts and a system 
of projectors to the transverse and traceless modes for the non-frame indices. 

In this context an important issue is the gauge symmetry origin of the unphysical modes. In the initial formulation there is no trace of such a symmetry. But, starting from the idea that such a symmetry exists but is hidden, we have shown that it can be unfolded by incorporating nonlocal terms in the theory. That is,  the hidden gauge symmetry can be made manifest in the action at the price of introducing non-local terms in it. One can turn around this conclusion by saying that non-locality is a gauge artifact, because it disappears if we return to the initial formulation of the theory. In other words, the HS-YM models can take a perturbatively local or a non-local form. In both cases they are characterized by a YM-like HS gauge symmetry, which may be the passe-partout for renormalization and unitarity. This is the bet of our research program.

\section*{Acknowledgements}

L.B. would like to thank Radu Constantinescu for inviting him to participate in this 2020 "Workshop on Quantum Fields and Nonlinear Phenomena". The research of S.G. has been supported by the Israel 
Science Foundation (ISF), grant No. 244/17.

\end{document}